          [1999/12/02 1.01 (PWD)]
\documentclass[a4paper,epsfig,twocolumn]{esapub} 
\usepackage{natbib}
\usepackage{graphicx}

\title{Nonlinear MHD Waves in the Solar Wind Plasma Structures}
\author{D. Tsiklauri,  V.M. Nakariakov and T.D. Arber}
\affil{Physics Department, University of Warwick, Coventry, 
   CV4 7AL, U.K., Email: tsikd@astro.warwick.ac.uk}

\begin{document}

\keywords{Magnetohydrodynamics(MHD)-- waves -- Sun: 
activity -- Sun: wind -- Sun: corona}

\maketitle

\begin{abstract}
 We investigate the interaction of magnetohydrodynamic waves with
plasma density  inhomogeneities. 
Our numerical study of the full MHD equations shows that: 
(A) Plasma density inhomogeneities are a source of
non-linear generation of transverse compressive waves by a plane
Alfv\'en wave, and
 substantially enhance
(by about a factor of 2) 
the generation of the longitudinal compressive
waves.
(B) Attained maximal values of the 
generated transverse compressive waves are  insensitive to
the strength of the plasma density inhomogeneity and the
initial amplitude of the Alfv\'en wave.
(C) Efficiency of the generation 
 depends weakly upon the plasma $\beta$ parameter. 
The maximum generated amplitude of transverse compressive wave,
up to 32 \% of the initial Alfv\'en wave amplitude,
is reached for about $\beta=0.5$.
The results obtained demonstrate that plasma inhomogeneities 
 enhance the efficiency of the non-linear wave
coupling.
\end{abstract}

\section{Introduction}

The structure of the solar corona and solar wind is
governed by the interaction between outward streaming
coronal plasma and the existing strong magnetic fields.
It is believed that the fast solar wind is associated
with  coronal holes within which the magnetic field
is unipolar, open and probably has further fine structuring.
The slow wind is believed to originate above and at the boundaries
of the coronal streamers where the magnetic fields are bipolar,
which forces the plasma to flow around regions where
the field strength is sufficient to slow down the flow.
The importance of Alfv\'en waves for the 
energetics of the corona and solar wind has been 
realized since their discovery \citep{tm}.
The interest in wave phenomena in the
solar wind is especially timely now, in the context
of interpretation of the CLUSTER observations.
The Ulysses observations already revealed a sharp boundary
between fast and slow wind. Besides, there are indications that
there is a further fine structuring of plumes in the coronal holes,
so called spaghetti.
Therefore, it is of prime importance to the study interactions
of the Alfv\'en waves with these and possibly other 
structures.
The weakly non-linear regime, relevant to  lower
coronal applications, has been studied in great detail in 
\citet{nrm97}, \citet{nrm98}, \citet{Botha}, \citet{td1}.
The relevance of this study is prompted by the observations
of large amplitude Alfv\'en waves in the solar wind.
In this work we simulate numerically the 
interaction of a {\it strongly non-linear} Alfv\'enic pulse 
with a one-dimensional, perpendicular to the magnetic field, 
plasma inhomogeneity.

\section{The model}

We use the equations of ideal MHD
$$
\rho {{\partial \vec V}\over{\partial t}} + 
\rho(\vec V \cdot \nabla) \vec V = - \nabla p -{{1}\over{4 \pi}}
\vec B \times {\rm curl} \vec B, \eqno(1)
$$
$$
{{\partial \vec B}\over{\partial t}}= {\rm curl} (\vec V \times \vec B),
\eqno(2)
$$
$$
{\rm div} \vec B =0, \eqno(3)
$$
$$
{{\partial \rho}\over{\partial t}} + {\rm div}(\rho \vec V)=0,
\eqno(4)
$$
where $\vec B$ is the magnetic field, $\vec V$ is plasma velocity,
$\rho$ is plasma mass density, and $p$ is plasma thermal pressure
for which the adiabatic variation law is assumed.
We solve equations (1)-(4) in Cartesian coordinates ($x,y,z$)
and for simplicity assume that there is no variation in the
$y$-direction, i.e. ($\partial / \partial y =0$). The equilibrium
state is taken to be an inhomogeneous plasma of density $\rho_0(x)$
and a uniform magnetic field $B_0$ in the $z$-direction.
We consider the physical situation when  {\it initially} longitudinal 
($V_z$, $B_z$) and transverse ($V_x$, $B_x$) compressive
perturbations are absent and the  initial amplitude of
the linearly polarized Alfv\'en wave ($V_y$, $B_y$) is essentially non-linear 
(relative amplitude normalized to the unperturbed value 
$A=0.5$). In this case,
if the Alfv\'en wave is  initially a plane wave, the subsequent
evolution of the wave, due to the difference in local Alfv\'en speed
across the $x$-coordinate, leads to the distortion of the wave front
(phase mixing).
Hence the appearance of transverse (with respect to the 
applied magnetic field) gradients, which grow in time as a power-law.
This transverse gradient is the only source  of the
transverse compressive perturbations, while longitudinal compressive
perturbations are also generated due to non-linear coupling
to the Alfv\'en wave.

\section{Numerical Results}

As in  \citet{Botha}, \citet{td1}, 
we have used the following background density profile
$$
\rho_0(x)=3-2 \, \tanh(\lambda x). \eqno(5)
$$
Here, $\lambda$ is a free parameter which controls
the steepness  of the density profile gradient. In our 
normalization, which is the same as that of \citet{Botha}, \citet{td1}, 
the local Alfv\'en speed is
$C_A(x)=1/ \sqrt{3-2 \, \tanh(\lambda x)}$.
\begin{figure}[]
\includegraphics[width=1.1\linewidth]{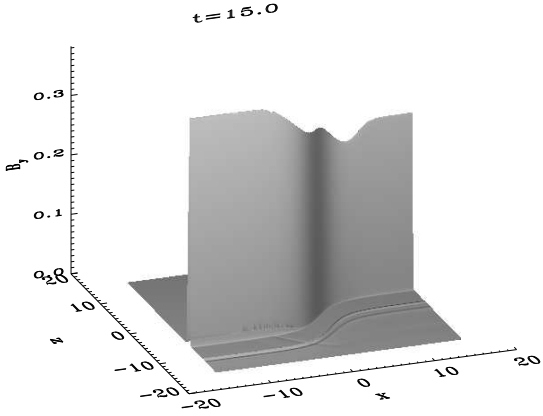}  
\caption{Snapshot of $B_y$ at $t=15.0$.
Here $A=0.5$, $\lambda=0.3125$, $\beta=2$.}
\end{figure}
Full numerical calculation of the non-linear set of MHD
equations (1)-(4), in the above mentioned geometry,
has been performed using {\it Lare2d} \citep{Arber}.
{\it Lare2d} is a numerical code which operates by taking
a Lagrangian predictor-corrector time step and after each
Lagrangian step all variables are conservatively
re-mapped back onto the original Eulerian grid
using Van Leer gradient limiters.
This code was also 
used to produce the results in  \citet{Botha}, \citet{td1}.
\begin{figure}[]
\includegraphics[width=1.1\linewidth]{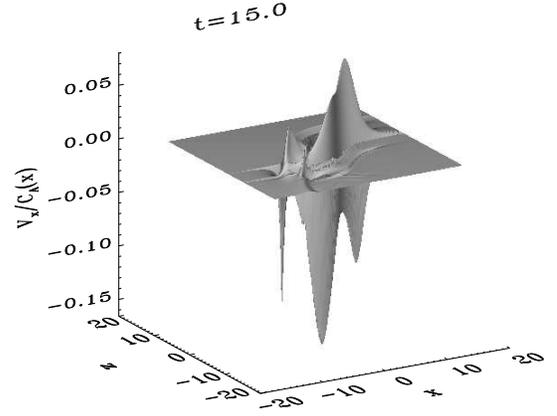}  
\caption{Snapshot of $V_x/C_A(x)$ at $t=15.0$.
Here $A=0.5$, $\lambda=0.3125$, $\beta=2$.}
\end{figure}
We set up the code in such a way that initially longitudinal 
and transverse  compressive
perturbations as well as density perturbation 
are absent and the  initial amplitude of
the Alfv\'en pulse is sufficiently non-linear, i.e.  A=0.5. 
In the numerical simulations
the Alfv\'en perturbation 
is  initially a plane (with respect to $x$-coordinate) 
pulse, which has a Gaussian structure in the $z$-coordinate,
and is moving at the local Alfv\'en speed $C_A(x)$:
$$
B_y(x,z,t)=A \exp \left(-{{(z-C_A(x)\, t)^2}\over{\delta}}\right).
\eqno(6)
$$
Here, $\delta$ is a dimensionless free parameter which controls the width
of the  initial Gaussian Alfv\'en pulse and it was fixed
at 0.1 in our numerical runs.
Initially, $V_y$ is non-zero too, and in our normalization
it is related to $B_y$ as
$V_y=-C_A(x)\, B_y$, while all other quantities are set equal to
zero.
The simulation box size is set by the limits $-15.0 < x < 15.0$
and  $-15.0 < z < 15.0$ and the pulse starts to move from
point $z=-12.5$ towards the positive $z$'s.
When $\lambda$-parameter was fixed at 0.125 we had to
increase our simulation box 1.5 times in order to accommodate
fully the density inhomogeneity. In this case we increased
the number of grid point by 1.5 in each direction
in order to preserve the same simulation resolution.
\begin{figure}[]
\includegraphics[width=1.1\linewidth]{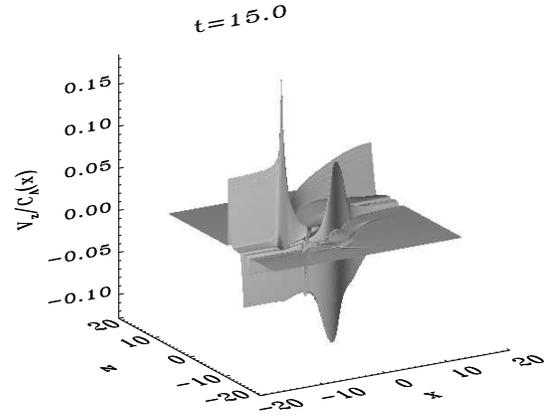}  
\caption{Snapshot of $V_z/C_A(x)$ at $t=15.0$.
Here $A=0.5$, $\lambda=0.3125$, $\beta=2$. }
\end{figure}
Fig. 1 presents a snapshot of $B_y$ at time $t=15$.
It can be seen from the graph that 
because of the difference in local Alfv\'en speeds
(Note, that for $x<0$ $C_A(x)=1/\sqrt{5}$ and 
for $x>0$ $C_A(x)=1$) the initially flat (with respect to $x$-coordinate)
Gaussian pulse has been distorted along $x=0$ axis.
This distortion of the pulse front creates a transverse
gradient ($\partial B_y / \partial x$) which 
is a driving force for the generation of
the transverse compressive wave ($V_x$).
Interesting observation is that leftmost ($B_y(-15,z,t)$)
and rightmost ($B_y(15,z,t)$) wings of the Alfv\'en pulse
decay due to shock formation at different rates.
Namely, the rightmost wing where local Alfv\'en velocity is larger
dissipates at a larger rate. This can be roughly explained by the
fact that in the scalar Cohen-Kursrud equation 
(cf. \citet{vnl}), which describes
the evolution of a weakly non-linear Alfv\'en wave until the
shock formation, the non-linear term is proportional to
$(C^3_A/(C^2_A-C^2_s))$, where $C_s$ is the speed of sound.
\begin{figure}[]
\includegraphics[width=1.1\linewidth]{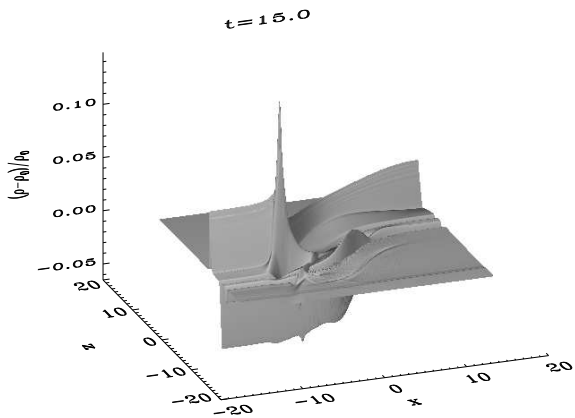}   
\caption{Snapshot of $(\rho-\rho_0)/ \rho_0$ at $t=10.0$.
Here $A=0.5$, $\lambda=0.3125$, $\beta=2$.  }
\end{figure}
Therefore it is expected that the shock formation would be 
reached quicker where the local Alfv\'en velocity is greater (in this
case rightmost wing). Yet another interesting feature
can be seen form this graph which is appearance of a small
bump in the solution at $x=0$ (phase-mixing region).
This could be explained by an energy exchange of the
Alfv\'en wave with the transverse and longitudinal compressive 
waves. The latter is expected to happen exactly
along the $x=0$ line where coupling between all three modes
is strongest.
We present solution of  Eqs.(1)-(4) with
the above described equilibrium and the initial conditions for the
case when plasma-$\beta$ is 2.
The $\lambda$-parameter was fixed at 0.3125 which corresponds
to the rather smooth maximum in the dependence of $max(|V_x(x,z,t)|)$
versus $\lambda$ for this value of $\beta$ 
(cf. Fig. 5 and pertaining discussion below).
\begin{figure}[]
\includegraphics[width=1.1\linewidth]{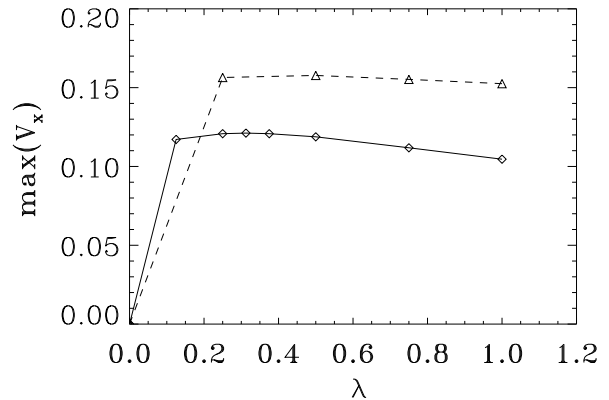}  
\caption{Dependence of $max(|V_x(x,z,t)|)$
versus $\lambda$ for $\beta=2$ (solid curve) and $\beta=0.5$ 
(dashed curve). Here, $A=0.5$. }
\end{figure}
At time $t=0$ transverse and longitudinal compressive 
waves as well as density perturbation were absent from the
system, while at time $t=15$ as we see from Figs. 2 and 3, 
these physical quantities grew up to a substantial fraction of the
initial Alfv\'en wave amplitude. 
The transverse compressive wave, $V_x$, is generated by the 
$B_y \partial B_y / \partial x$ driving term see for details
\citet{td1}. In the homogeneous plasma ($\lambda=0$)
for a plane wave
this term is identically zero, thus,
in the numerical runs with $\lambda=0$
we have seen no generation of $V_x$ at all.
This is the plasma inhomogeneity that leads to the
generation of the transverse compressive wave in this case. 
The situation with the longitudinal compressive wave, $V_z$,
is different: even in the absence of the plasma density inhomogeneity
the efficient generation of $V_z$ still occurs due to the
$B_y \partial B_y / \partial z$ term. 
By looking at Fig. 3 we gather that the solution consists of
two parts: first, two positive and negative spikes, which dominate
the the solution by absolute value, which indeed 
come from the phase-mixing
region, i.e. part of the solution which is caused by density inhomogeneity
and the second, relatively
 plane wave front with a substantial deep in the
middle which is caused by the non-linear coupling to the Alfv\'en wave.
It can be also gathered that the second part is moving
at about twice of the local Alfv\'en speed because $\beta=2$.
\begin{figure}[]
\includegraphics[width=1.1\linewidth]{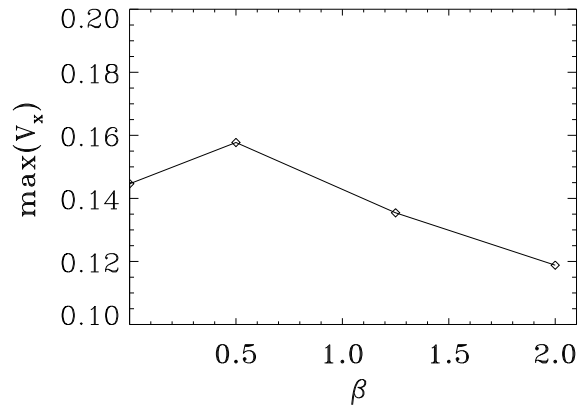}  
\caption{Dependence of $max(|V_x(x,z,t)|)$
versus $\beta$. Here, $\lambda=0.5$ and $A=0.5$. }
\end{figure}
The relative density perturbation presented in Fig. 4 has similar structure
as $V_z$, but it is enhanced in the region of  interaction.
$max(|(\rho-\rho_0)/ \rho_0|)=0.15$
in this case, i.e. 30 \% of the amplitude of initial Alfv\'en perturbation.
Next, we explore parametric space of the
problem. In particular, we investigate how the
maximal value of the generated transverse compressive wave
depends on the plasma density inhomogeneity parameter, $\lambda$,
plasma $\beta$, and initial amplitude of the Alfv\'en wave, $A$.
In Fig. 5 we plot dependence of $max(|V_x(x,z,t)|)$
versus $\lambda$ for $\beta=2$ (solid curve) and $\beta=0.5$ 
(dashed curve). There are two noteworthy features in this graph.
First, the maximal value of the generated transverse compressive wave
depends on the plasma density inhomogeneity parameter rather
weakly, and second, the efficiency of the generation of $V_x$
is somewhat larger in the $\beta=0.5$ case than in the
case of $\beta=2$. In order to explore the latter feature
further, we made 4 runs of {\it Lare2d} code for the
different values of $\beta$, while $\lambda$ was fixed at
0.5. The results are presented in Fig. 6. We gather from the
graph that there is a given value of $\beta$ for which 
a  maximal value of $max(|V_x(x,z,t)|)$
is observed.
\begin{figure}[]
\includegraphics[width=1.1\linewidth]{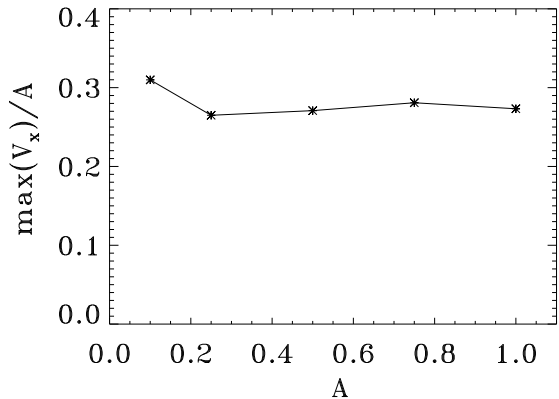}  
\caption{Dependence of $max(|V_x(x,z,t)|)/A$
versus $A$. Here, $\lambda=0.5$ and $\beta=1.25$. }
\end{figure}
Yet another valuable insight can be obtained by studying
the dependence of the maximal value of  generated transverse 
compressive wave on the initial amplitude of the Alfv\'en wave
as our problem is essentially non-linear.
In Fig. 7 we plot results of 4 numerical runs for different
values of $A$, while $\lambda$ and $\beta$ where fixed at
0.5 and 1.25 respectively.
We gather from Fig. 7 that quite unexpectedly the ratio
$max(|V_x(x,z,t)|)/A$ is rather insensitive to the variation of the
initial amplitude of the Alfv\'en wave, and it varies around about
0.27. In other words, irrespective of $A$, the
generated transverse compressive wave attains about 27 \% of the
initial amplitude of the Alfv\'en wave for this value of $\beta$.
Answering the question what determines the observed saturation
level of maximal value of the generated transverse compressive wave,
we conjecture that, based on the fact that the saturation level
is rather insensitive to the variation of $\lambda$ (see Fig. 5)
and it somewhat stronger depends of the variation of $\beta$ 
(see Figs. 5 and 6), it is mostly dictated by the given
value of $\beta$.
Based on the information provided by Fig. 6 we conclude
that {\it the most efficient generation of the transverse compressive wave
occurs when $\beta=0.5$ and, irrespective of $\lambda$ and $A$
parameters,
transverse compressive wave attains about 32 \% of the 
initial amplitude of the Alfv\'en wave}.

\section{Conclusions}

Our numerical study of the full MHD equations revealed  the following:
(A) Plasma density inhomogeneity is a source of
non-linear generation of transverse compressive waves by a plane
Alfv\'en wave, and
it substantially enhances
(by about a factor of 2) 
the generation of the longitudinal compressive
waves.
(B) Attained maximal values of the 
generated transverse compressive waves are  insensitive to
the strength of the plasma density inhomogeneity and the
initial amplitude of the Alfv\'en wave.
(C) Efficiency of the generation 
 depends weakly upon the plasma $\beta$ parameter. 
The maximum generated amplitude of transverse compressive wave,
up to 32 \% of the initial Alfv\'en wave amplitude,
is reached for about $\beta=0.5$.
(D) Inhomogeneity of the medium is essential for the
generation of weakly compressible MHD turbulence.
The results obtained demonstrate that plasma inhomogeneities 
 enhance the efficiency of the non-linear wave
coupling.

It is worthwhile to mention that in this work we neglected
variation in the
$y$-direction, i.e. we set $\partial / \partial y =0$.
This, in turn, prevents linear coupling of the Alfv\'en waves
to the compressive waves. However, validity of 
this assumption cannot be tested observationally so far.
Yet another interesting possibility of the generation of 
compressive waves exists even in the homogeneous plasma when
either Alfv\'en wave front is not {\it initially flat} with respect
to $x$-coordinate or initially flat Alfv\'en wave travels
over a {\it spatially confined} plasma density inhomogeneity.
In both cases we should observe effective generation of the
the compressive waves which possibly do not experience
suppression caused by the fact that $\partial / \partial x \to \infty$
as in the case considered here and in \citet{td1}.

{\bf Acknowledgments:} 
Numerical calculations of this work were
done using the PPARC funded Compaq MHD Cluster in St. Andrews.

\end{document}